\documentclass[%
twocolumn,superscriptaddress,
aps, prl,
]{revtex4-2}
\usepackage[utf8]{inputenc}

\usepackage{graphicx}
\usepackage{dcolumn}
\usepackage{bm}
\usepackage{blindtext}
\usepackage{xcolor}
\usepackage{xr}
\usepackage{amsmath, amsfonts, amssymb}
\usepackage{stmaryrd} 
\usepackage[mathlines]{lineno}

\makeatletter
\newcommand*{\addFileDependency}[1]{
  \typeout{(#1)}
  \@addtofilelist{#1}
  \IfFileExists{#1}{}{\typeout{No file #1.}}
}
\makeatother

\newcommand*{\myexternaldocument}[1]{%
    \externaldocument{#1}%
    \addFileDependency{#1.tex}%
    \addFileDependency{#1.aux}%
}
\myexternaldocument{SI} 

\begin{document}

\preprint{APS/123-QED}

\title{Interfacial instability of confined 3D active droplets}%
\date{\today}
\author{Bennett C. Sessa}
\thanks{These authors contributed equally.}
\affiliation{Department of Physics, Brandeis University, Waltham, Massachusetts 02453, USA}

\author{Federico Cao}
\thanks{These authors contributed equally.}
\affiliation{School of Engineering, Brown University, Providence, RI 02912, USA.}
\affiliation{Center for Fluid Mechanics, Brown University, Providence, RI 02912, USA.}

\author{Robert A. Pelcovits}
\affiliation{Department of Physics, Brown University, Providence, RI 02912, USA.}
\affiliation{Brown Theoretical Physics Center, Brown University, Providence, RI 02912, USA.}

\author{Thomas R. Powers}
\affiliation{School of Engineering, Brown University, Providence, RI 02912, USA.}
\affiliation{Center for Fluid Mechanics, Brown University, Providence, RI 02912, USA.}
\affiliation{Department of Physics, Brown University, Providence, RI 02912, USA.}
\affiliation{Brown Theoretical Physics Center, Brown University, Providence, RI 02912, USA.}

\author{Guillaume Duclos}
\homepage{gduclos@brandeis.edu}
\affiliation{Department of Physics, Brandeis University, Waltham, Massachusetts 02453, USA}

\date{\today}
\begin{abstract}

Instabilities of fluid--fluid interfaces are ubiquitous in passive soft matter. Adding activity to the interface or either fluid can dramatically change the stability of the interface. Using experiment and theory, we investigate the interfacial instability of a deformable 3D active nematic liquid crystal droplet in the isotropic phase surrounded by a passive fluid and confined between two parallel plates. Spontaneous active flows drive the growth of undulations along the active/passive interface, with the mode number of the fastest--growing mode increasing with droplet radius and decreasing with gap height. We apply the lubrication approximation to a minimal nematohydrodynamic model to determine the growth rates of all interfacial modes. The magnitude of the growth rate is determined by the active timescale and the relaxation timescales associated with liquid crystalline order as well as capillary and viscous stresses. We find multiple points of agreement between experiment and theory, including the shape evolution of individual droplets, the growth rates of unstable modes averaged across many droplets, and the extensional shear flows observed within droplets.

\end{abstract}

\maketitle
Active fluids are composed of energy--consuming constituents that can generate spontaneous flow~\cite{Marchetti2013,Ramaswamy2019,DellArciprete2018, Riedel2005, Sanchez2012, Dunkel2013}. Unconfined, they are often unstable~\cite{Simha2002, Marenduzzo2007,Miles2019}. In extensile active nematics, confinement selects the most unstable wavelength~\cite{Chandrakar2020,Strbing2020}, and strong confinement can even suppress flows, leading to a quiescent active phase~\cite{Voituriez2005,Duclos2018,Alam2024}. The stability of an active fluid confined by a deformable boundary, such as an interface or membrane, is markedly different because the boundary can deform in response to the stress generated by the active constitutive particles. The interplay of active and restoring surface forces underlies interfacial instabilities in many experimental systems, such as synthetic colloids, biofilms, and cells~\cite{Soni2019_odd, Ganesh2025, MartnezCalvo2022, BarZiv1999,Alert2019}. However, the instability of a deformable interface between an active liquid crystal and a passive fluid is still largely unexplored, especially in 3D. Experimental realizations of interfacial deformations in cytoskeletal active matter are rare because the active stress is usually several orders of magnitude smaller than the restoring stresses controlled by surface tension~\cite{Foster2023, Adkins2022}. Activity--driven deformations have been observed in phase--separated polymeric fluids~\cite{Adkins2022,Zhao2021} whose surface tension is almost six orders of magnitude smaller than the oil/water interface conventionally used to confine active nematics in droplets~\cite{Sanchez2012,Chen2021,Alam2024}. Previous theoretical work elucidates the antagonistic roles that activity and capillary forces play on the stability of active interfaces~\cite{SalbreuxJulicher2017, Nagilla2018, Soni2019, Dhar2025,Blow2014}. Below a critical activity, surface tension stabilizes all perturbations. Above this activity, a finite wavelength perturbation grows, and the fastest--growing mode increases with activity and droplet size~\cite{Alert2022, Nagilla2018, Soni2019}. Most theoretical studies on deformable active nematic droplets are 2D~\cite{Blow2014, Giomi2014, Soni2019, Dhar2025, Nagilla2018}, but characterizing the interfacial instability of experimental active nematic droplets requires a consideration of their 3D geometry. While some studies have considered the nonlinear evolution of interfaces in 3D~\cite{Ruske2021,Adkins2022, Zhao2024,Sciortino2025}, we focus on the onset of the instability.
\begin{figure*}
    \centering
    \includegraphics[width=\linewidth]{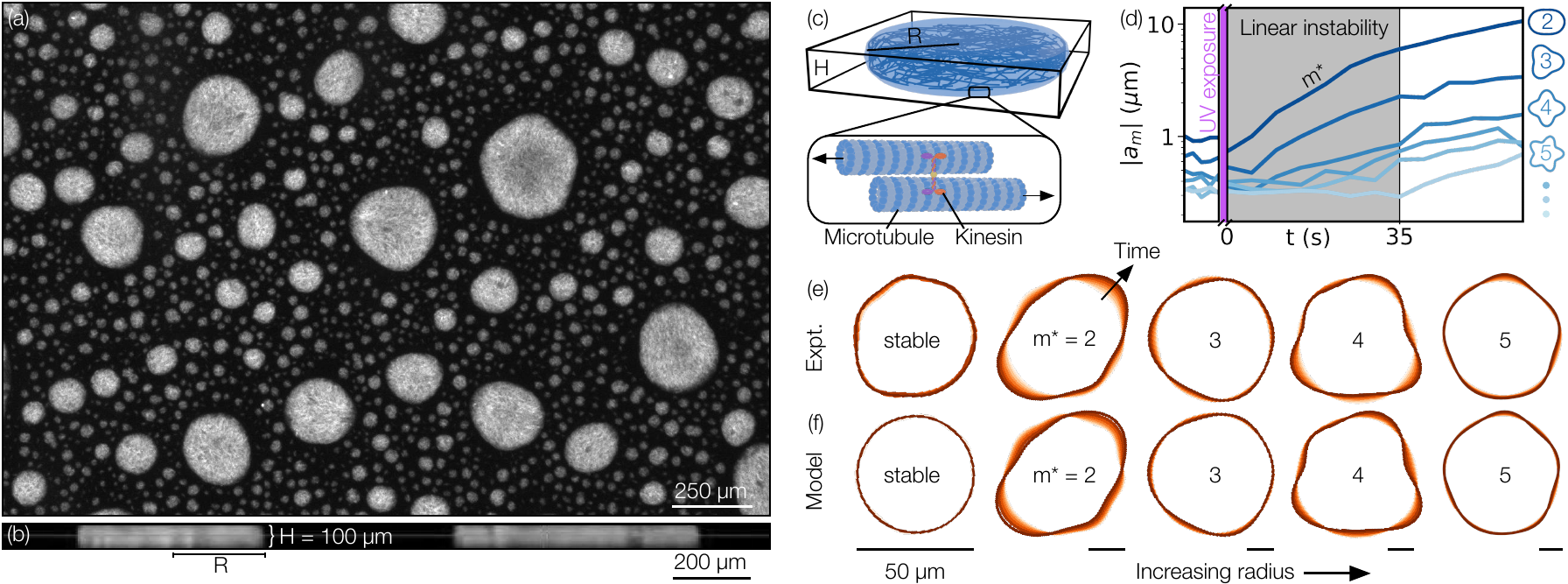}
    \caption{Interfacial instability of active droplets confined between two parallel plates. (a) Top view of experimental active droplets (Microtubules are fluorescent). (b)  Side view of confined droplets. (c) Schematic of an active droplet containing extensile microtubule bundles and molecular motors. (d) Log--linear plot of time evolution of the radial mode amplitudes $a_m$. We shine UV light to uncage the ATP. (e) Time evolution of droplet interfaces experimentally and (f) theoretically after fitting the model parameters and retaining the first eight radial modes. Experimental and theoretical droplets have the same number of undulations with similar orientations. Color shows the time evolution from orange to red. Scale bars are all $30\, \mathrm{\mu m}$, and droplet radius increases from left to right. Gap height $H=20\,\mathrm{\mu m }$ for (a,d--f) and $H=100\, \mathrm{\mu m}$ for (b).} 
    \label{fig:1}
\end{figure*}

In this Letter, we combine experiments and theory to investigate the interfacial instability of 3D active droplets in a passive viscous fluid confined between two narrowly separated parallel plates, a geometry known as a Hele--Shaw cell [Fig.~\ref{fig:1}(a--c)]. The active fluid is a 3D extensile active nematic liquid crystal in the isotropic phase. We apply a lubrication approximation to a minimal nematohydrodynamic model and find that the balance of active and restoring forces determines the most unstable mode. Active hydrodynamic flows deform the low surface tension interface and equally excite all the radial interfacial modes. The passive restoring forces from surface tension and viscosity stabilize the interface. Due to the thin gap geometry, changing the droplet radius and gap height have opposite effects on the instability. Increasing droplet radius reduces the Laplace pressure, favoring the growth of high mode number (short wavelength) undulations. In contrast, increasing the gap height reduces viscous forces, favoring the growth of low mode number (long wavelength) undulations.

\textit{Experiments:} We study a 3D isotropic active liquid crystal composed of microtubules and ATP--powered kinesin molecular motors~\cite{Sanchez2012} [Fig.~\ref{fig:1}(c)] embedded within a phase--separated polymeric fluid of PEG and Dextran~\cite{Adkins2022}. The microtubules and the motors form extensile bundles that partition within the Dextran--rich phase, forming active droplets of various sizes surrounded by a passive PEG--rich viscous fluid [Fig.~\ref{fig:1}(a)]. The binary fluid is confined between two parallel plates that have been treated with a PEG brush to prevent wetting of the Dextran--rich active droplets [Fig.~\ref{fig:1}(b)]. Uncaging ATP with UV light triggers a transition from passive quiescent droplets to active droplets (Movie S1). The extensile bundles form a continuously reconfiguring network that advects the surrounding fluid, powering spontaneous collective flows. It is important to note several features that differentiate these 3D active droplets from the well--established active nematic shells where kinesin--microtubule bundles are depleted at a 2D oil/water interface~\cite{Sanchez2012,keber2014,Guillamat2018,Sciortino2025}. First, in our experiments, the extensile bundles are dispersed in 3D, completely filling the droplet. Consequently, active flows are not restricted to the plane of the interface, and stresses normal to the interface need to be considered. Second, the microtubule bundles are dilute ($\sim$ 0.1\% volume fraction) forming a 3D active nematic liquid crystal in the isotropic phase instead of a dense 2D active nematic. Finally, as mentioned earlier, the surface tension of the PEG/Dextran interface is almost six orders of magnitude lower than that of the oil/water interface~\cite{Adkins2022}.

We observe that after activation of the molecular motors, many small droplets do not change shape, whereas larger droplets systematically do, a phenomenon reminiscent of a morphological interfacial instability. As the droplet size increases, the number of ripples on the interface increases [Fig.~\ref{fig:1}(e)], in agreement with a hydrodynamic model presented below [Fig.~\ref{fig:1}(f)]. We decompose the interface into radial modes and find that the amplitude of the modes grows exponentially, with the fastest--growing mode corresponding to the number of ripples observed [Fig.~\ref{fig:1}(d)]. The exponential growth is followed by a nonlinear regime during which interfacial ripples continuously grow and decay over time.

\begin{figure}[!t]
    \centering
    \includegraphics[width=\linewidth]{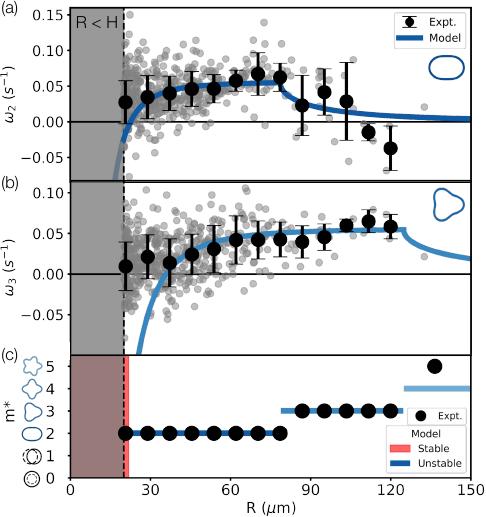}
    \caption{Fastest--growing mode number increases with droplet radius for $H = 20 \, \mathrm{\mu m}$. Measured growth rates (disks) and model growth rates (lines) for the (a) second and (b) third radial modes.  The model parameters are determined by fitting only $\omega_2$. For $\omega_3$, the model prediction is obtained without adjusting these parameters. Error bars represent the standard deviation of the binned growth rates. (c) Fastest--growing mode number $m^*$ increases with droplet radius $R$.}
    \label{fig:radius} 
\end{figure}

To be more quantitative, we measure the growth rates of the first ten modes of the interface for hundreds of droplets of varying radii and fixed thickness. We note that for a given droplet radius, the growth rates of the second harmonic are widely distributed with some positive and some negative, corresponding to unstable and stable droplets, respectively. For the smallest experimentally accessible radii, we observe a mixture of stable and unstable droplets. We cannot statistically measure a critical radius below which experimentally accessible droplets are stable. Still, we observe that the proportion of stable droplets decreases as their radius increases [Fig.~S10]. Binning similarly sized droplets together reveals that $\omega_2$, the average growth rate of the second radial harmonic, initially increases and then decreases with droplet radius [Fig.~\ref{fig:radius}(a)]. The average growth rate of the third radial harmonic, $\omega_3$, also increases with radius [Fig.~\ref{fig:radius}(b)] and becomes larger than $\omega_2$ for $R \gtrsim 80\,\mu$m [Fig.~\ref{fig:radius}(c)]. Some of the widest droplets achievable experimentally exhibit growth of the fourth and fifth radial modes [Fig.~\ref{fig:1}(e), Fig.~\ref{fig:radius}(c), Fig.~S9].
\begin{figure*}[!t]
    \centering
    \includegraphics[width=2\columnwidth]{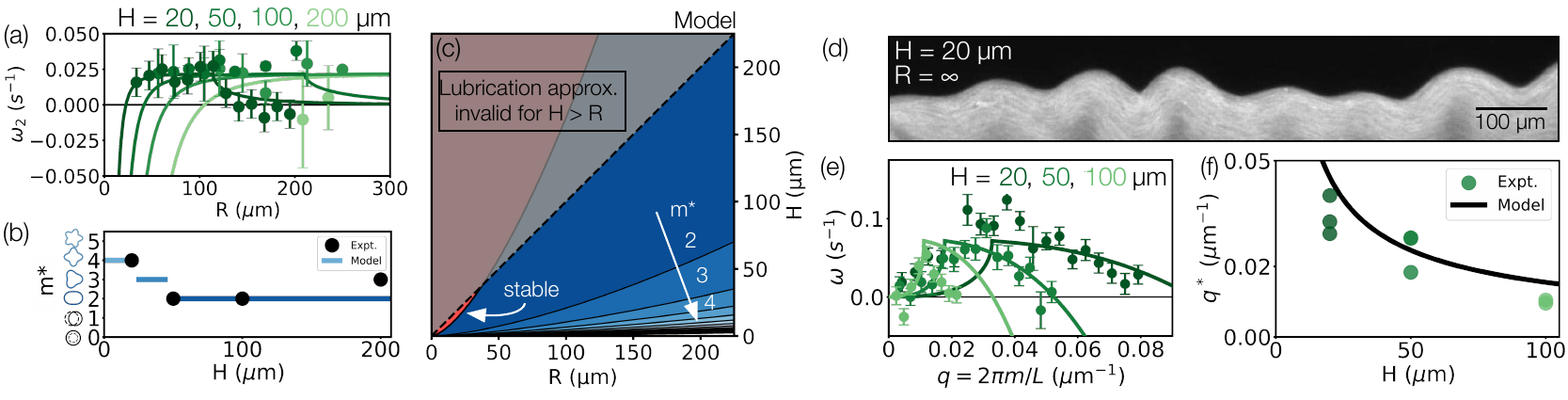}  
    \caption{Fastest--growing mode number $m^*$ and wavenumber $q^* = 2\pi m^*/L$ decrease with gap height $H$. (a) Measured growth rates (disks) and model growth rates (lines) for the second radial harmonic at different gap heights. The model parameters are determined by fitting only $\omega_2$ for $H=20\,\mu$m. (b) Fastest--growing mode number $m^*$ decreases with increasing gap height $H$ for $R=200\,\mu$m. (c) Phase diagram of the fastest--growing mode number from the model using the parameters in (a). (d) Instability at an initially flat active/passive interface. (e) Dispersion relation for a flat interface of length $L$ at different gap heights. The fit parameters are determined by fitting $\omega$ only for $H=50\,\mu$m. (f) Model predicts that the fastest--growing wavenumber $q^*$ for the initially flat interface decreases with gap height as $q^*\propto  H^{-2/3}$.}
    \label{fig:height}
\end{figure*}

We find that increasing the thickness of the confined droplet has the opposite effect as increasing the droplet radius. We measure the growth rates of the radial modes for droplets of varying radii and four different gap heights ranging from 20 to 200$\,\mu$m. The droplet radius necessary to achieve a given growth rate must increase with increasing droplet thickness [Fig.~\ref{fig:height}(a)]. At a fixed radius, the fastest--growing mode number decreases with gap height. In summary, thin and wide droplets ($R\gg H$) have higher fastest--growing mode numbers compared to thick and narrow droplets ($R\sim H$) which have lower fastest--growing mode numbers and tend to be stable.

\textit{Model:} To understand our experimental observations, we develop a minimal hydrodynamic model. The experiments occur at low Reynolds number because of the small spatial dimensions and slow flows (Re $\sim 10^{-8}$, droplet size $\sim 100\,\mu$m, flow speed $\sim 1\,\mu$m/s, viscosity  $\sim 10 \, \mathrm{mPa \cdot s}$). The passive outer fluid is described using the Stokes equations. The active fluid is described using a modified Stokes equation: we use a minimal nematohydrodynamic model for an incompressible extensile active nematic liquid crystal in the isotropic phase~\cite{Varghese2020}. The microtubule bundles are modeled by active force dipoles, which are represented by the divergence of the nematic order parameter tensor $\mathsf{Q}$ in this continuum model~\cite{Hatwalne2004}. The equations are linearized about a state of no flow and isotropic order, corresponding to the state of the experimental droplets before the activity is turned on. The liquid crystalline order emerges from a balance between a tendency to relax to isotropy and align due to activity--driven shear flows [Eq.~\eqref{eq:Q}]~\cite{DeGennes1969, Varghese2020, Chandrakar2020}.

A distinguishing feature of this work is the application of the lubrication approximation~\cite{Bensimon1986} to account for the fact that the droplets are wide and thin [Fig.~\ref{fig:1}(b)]. In this lubrication approximation, gradients in the velocity and the order parameter tensor across the gap are greater than gradients along the plane. For simplicity, we neglect Frank elasticity because including it does not substantially change our results [Fig.~S6]. Therefore, there are no boundary conditions enforced on $\mathsf{Q}$. We further assume no--slip boundary conditions on the velocity field at the top and bottom plates. The resulting equations for the active droplet are (see SI Sec.~1)
\begin{align}
    - \partial_{i} p + \eta \partial_z^2 u_{i} - \eta \tau_{\mathrm{a}}^{-1}\partial_z Q_{i z} &= 0 \\
    \left(\tau_\mathrm{LC} \partial_t+ 1\right)Q_{i z} - \lambda \tau_\mathrm{LC}\partial_z u_{i} &= 0 \label{eq:Q}\\
    \partial_z p &= 0,
\end{align}
where $p$ is the pressure, $u_i$ is the in--plane velocity ($i=x,y$), $Q_{iz}$ is the $iz$--component of the order parameter tensor, $\lambda \sim 1$ is the flow--tumbling parameter for long slender rods, and $\eta$ is the shear viscosity of the active and passive fluids (assumed equal for simplicity). We define two timescales for the liquid crystal. First, the active timescale $\tau_{a}=\eta/a$ decreases when activity $a$ increases and is proportional to velocity autocorrelation time~\cite{Chaitanya2022}. Second, the liquid crystal timescale $\tau_\mathrm{LC}$ characterizes the relaxation of the order parameter to isotropy. The active/passive interface has the following boundary conditions: the dynamic boundary condition where the jump in normal stress is given by the Laplace pressure, the kinematic boundary condition where the interface moves with the fluid, and continuity of the normal velocity.

To study the linear instability of the interface, we assume that the velocity, pressure, and order parameter tensor have time dependence $\exp(\omega t)$. Due to our linearization about the isotropic and quiescent state, and the neglect of Frank elasticity, we can eliminate the order parameter tensor from these equations which leads to Stokes equations for the velocity with an effective viscosity that depends on the activity (see SI Sec.~1):     
\begin{align}
   0 &= - \partial_{i} p +\eta_\mathrm{eff} \partial_z^2 u_{i}, \label{eq:modstokes} \\
    \dfrac{\eta_\mathrm{eff}}{\eta} &=1-\lambda\dfrac{\tau_\mathrm{a}^{-1}}{\omega+\tau_{\mathrm{LC}}^{-1}}. \label{eq:eta_eff}
\end{align}
Note that for small $\omega$, the effective viscosity decreases with increasing activity, consistent with rheological experiments on microtubule--based active gels~\cite{Gagnon2020,Marenduzzo2007_2}. When the real part of the effective viscosity is sufficiently negative, there is no resistance to spontaneous shear flow, and active flows drive the deformation of the droplet.


To understand the various timescales at play in the active instability, let us first consider the relaxation of a passive droplet slightly perturbed from its equilibrium shape, with radius $r(\theta,t)=R+h(\theta,t)$, where $r$ and $\theta$ are polar coordinates. We use simple scaling arguments to estimate how the relaxation time $\tau_{\mathrm{HS}}$ depends on the droplet radius $R$ and the gap height $H$. The balance of the forces arising from the pressure $p$ and the surface tension $\gamma$ of the interface leads to $p\sim \gamma h/R^2$. The pressure gradient drives a viscous flow, $p/R\sim (\eta /\tau_{\mathrm{HS}})( h/H^2)$. Thus, $1/\tau_{\mathrm{HS}}\sim \gamma H^2/(\eta R^3)$. Hence, for passive droplets, the relaxation is slower for larger droplets since the driving force---the pressure gradient---is smaller for larger droplets. In contrast, the relaxation is faster for larger gap heights $H$ since the viscous resistance is smaller.

Writing $h$ in a Fourier series $h \propto \cos(m\theta)$, where $m$ is the mode number, reveals how the growth rate of each mode depends on $m$,
\begin{equation}
    \omega_m = -\dfrac{\eta}{12\tau_{\mathrm{HS}} }\dfrac{m(m^2-1)}{\eta_{\mathrm{eff}}+\eta} ,\label{eq:dispersion}
\end{equation}
just as in the case of two passive fluids with two viscosities $\eta_{\mathrm{eff}}$ and $\eta$ \cite{Bensimon1986}.
For a passive droplet, all modes decay with $\omega_m < 0$. For an active droplet,
$\eta_{\mathrm{eff}}$ is a function of $\omega$ [Eq.~\eqref{eq:eta_eff}], which may be complex \footnote{Complex growth rates imply the existence of travelling waves \cite{Soni2019}. The study of these waves is out of the scope of this letter. Travelling waves have been observed in the nonlinear regime \cite{Adkins2022}}. Therefore, the real part of the growth rate is positive when the real part $\Re({\eta_{\mathrm{eff}} + \eta}) < 0$. As expected, the growth rate vanishes for the dilational mode $m=0$ due to incompressibility and for the translational mode $m=1$. For an active droplet, Eq.~\eqref{eq:dispersion} is an implicit equation for the dispersion relation. Solving for the growth rates, we find
\begin{multline}
    \omega^{\pm}_m = \dfrac{\lambda}{4 \tau_\mathrm{a}} -\dfrac{1}{2\tau_\mathrm{LC}} - \dfrac{M}{2 \tau_\mathrm{HS}} \\ 
    \pm \sqrt{ \left(\dfrac{\lambda}{4 \tau_\mathrm{a}} -  \dfrac{1}{2 \tau_\mathrm{LC}} - \dfrac{M}{2\tau_\mathrm{HS}}   \right)^2 -  \dfrac{M }{\tau_\mathrm{HS}\tau_\mathrm{LC}}  }, \label{eq:growth_rate}
\end{multline}
where $M = m(m^2-1)/24$. When the activity vanishes, the two relaxation modes associated with the interface and the liquid crystal are decoupled [Fig.~S5(a)]. The liquid crystal relaxes to isotropy without accompanying flows and the motion of the relaxing interface drives a flow. For non--zero activity, these two modes are coupled: flows drive order and order drives flows [Fig.~S5(b)]. Above a critical activity, the normal component of the spontaneous flows drives the interfacial instability [Fig.~S5(c)]. For a fixed height, there is a critical radius $R_c$ above which the interface is unstable [Fig.~\ref{fig:height}(c)], 
\begin{equation}
    R_c = H^{2/3} \left(\dfrac{\gamma}{\eta} \dfrac{2M \tau_{LC}\tau_a}{\lambda \tau_{LC}-2\tau_a} \right)^{1/3}.
\end{equation}
Similarly, for a fixed radius, there is a critical height below which the interface is unstable [Fig.~\ref{fig:height}(c)].

We find multiple points of agreement between the experiment and model. First, the model quantitatively captures how the mean growth rates of unstable droplets depends on their 3D geometry. When analyzing the effect of droplet radius, we fit the theoretical growth rate of $\omega_2$ [Eq.~\eqref{eq:growth_rate}] to the experimental data [Fig.~\ref{fig:radius}(a)] to determine the parameters $\tau_{\mathrm{a}}, \tau_{\mathrm{LC}},$ and $\tau_{\mathrm{HS}}$. Without modifying these parameters, we find quantitative agreement for the third radial mode [Fig.~\ref{fig:radius}(b)] and the fastest--growing mode [Fig.~\ref{fig:radius}(c)] as a function of droplet radius. Repeating this procedure for varying gap heights, we again find good agreement for the second radial mode [Fig.~\ref{fig:height}(a)] and the fastest--growing mode [Fig.~\ref{fig:height}(b)] as a function of droplet radius and gap height. Fig.~\ref{fig:height}(c) summarizes the opposing contribution of the gap height $H$ and the droplet radius $R$ to the interfacial instability. Second, the model predicts a similar evolution of the shape of individual droplets when starting from the experimental initial conditions for the amplitude and phase of each mode [Fig.~\ref{fig:1}(e,f), Fig.~S3]. The three time scales are $\tau_\mathrm{a}\sim 0.5\,$s, $\tau_\mathrm{LC}\sim 1\,$s, and $\tau_\mathrm{HS} \sim 1$--$1000\,$s. Our fit value for $\tau_\mathrm{a}$ yields an active stress $a\sim 50\,$mPa, which is consistent with previous measurements of the active stress at a flat quasi--2D active/passive interface \cite{Adkins2022}.

We further demonstrate agreement between our model and experiments by considering the limit of a 2D interface between an active and passive fluid confined between two parallel plates [Fig.~\ref{fig:height}(d), Movie S2]. We observe the growth of an interfacial instability whose fastest--growing wavenumber $q^*$ is finite [Fig.~\ref{fig:height}(e)] and decreases with increasing gap height [Fig.~\ref{fig:height}(f)]. We fit the theoretical growth rates for $H=50\,\mu$m and find good agreement between the model and the experimental dispersion relations at different gap heights [Fig.~\ref{fig:height}(e)]. Moreover, the model predicts that the fastest--growing wavenumber of a flat 2D active/passive interface decreases with gap height, which we confirm experimentally [Fig.~\ref{fig:height}(f)]: $q^* \propto H^{-2/3}$ (see SI Sec.~1). 

\begin{figure}[!t]
    \centering
    \includegraphics[width=\columnwidth]{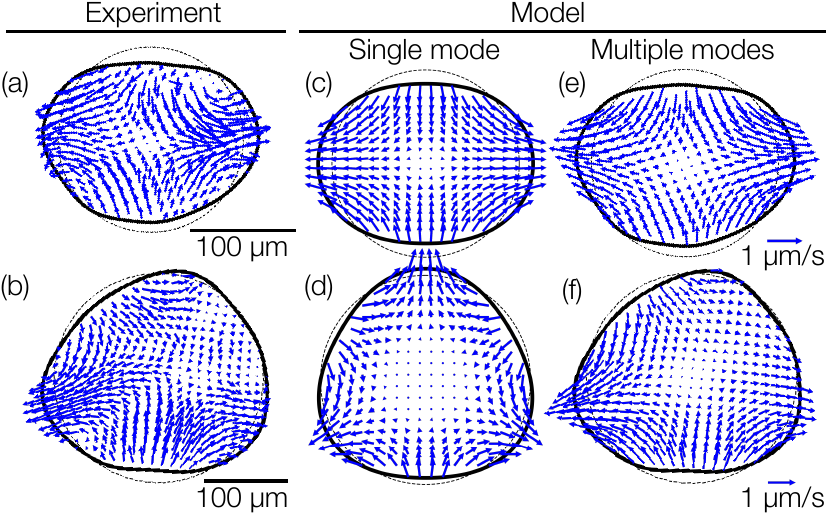}
    \caption{Extensional shear flow drives interfacial instability ($H=50\, \mathrm{\mu m}$). In the experiment, interfaces exhibiting mainly the second (a) and third (b) radial modes have predominantly dipolar and tripolar extensional shear flow (blue arrows) inside, respectively. (c,d) The model with a single extensional shear does not fully characterize the irregularities of the experimental active flows. (e,f) A superposition of the extensional shear modes better capture these features.}
    \label{fig:flow}
\end{figure}

\emph{Active flows.} So far, our investigation has focused solely on interfacial dynamics. Experiments and theory can also probe the nature of active flows driving the interfacial instability. Using particle--image velocimetry , we quantify the active flows inside unstable droplets and find that $n$--fold symmetric extensional shear flows drive the interfacial instability. Experimental droplets growing modes two and three exhibit predominantly two--fold and three--fold symmetric shear flows, respectively [Fig.~\ref{fig:flow}(a,b)]. Since a single extensional shear mode [Fig.~\ref{fig:flow}(c,d)] does not completely capture the minor asymmetries present in the experimental flows, we consider a superposition of the different extensional shear modes [Fig.~\ref{fig:flow}(e,f)]. Using the kinematic boundary condition, we establish a linear relationship between the Fourier amplitudes for the flows and the interface [SI Sec.~1: Eq.~(28), (30)]. Using the experimental droplets' interface and the fitted parameters, we reconstruct the instantaneous flows fields for individual droplets with fastest--growing modes $m^*=2$ and $m^*=3$ [Fig.~\ref{fig:flow}(e,f), Fig.~S2]. The quantitative agreement between the experiments and our minimal model confirms that the lubrication approximation captures the essential physics underlying the interfacial instability.

\emph{Discussion \& Conclusion.} In summary, above a critical radius or below a critical thickness, 3D droplets of shear--aligning isotropic active liquid crystal confined between two parallel plates are unstable. The interface grows a superposition of multiple modes, with the dominant one corresponding to the fastest--growing mode and with the same symmetry as the observed active shear flows.

Despite the similarities of the confined geometry and the growth of interfacial radial modes, we find the active instability is fundamentally distinct from the Saffman--Taylor instability where a less viscous fluid displaces a more viscous fluid. First, in the Saffman--Taylor instability, any external non--zero pressure gradient leads to the fingering instability. Comparatively, the active fluid is driven out of equilibrium internally at the microscopic level. The active interfacial instability occurs only when the activity exceeds a critical (non--zero) value. Second, the active instability occurs regardless of whether we consider an active droplet surrounded by a passive fluid or a passive droplet surrounded by an active fluid (Movie S3). In contrast, in the Saffman--Taylor instability, the interface is unstable only when the less viscous fluid invades the more viscous fluid. The reverse situation wherein the more viscous fluid invades the less viscous fluid is stable. Finally, the Saffman--Taylor instability is a quasi--2D instability that requires confinement; the deformations vanish for an unconfined 3D droplet or a flat 2D interface. In contrast, 3D active droplets and 2D active/passive interfaces can become unstable even without the Hele-Shaw confining geometry \cite{Zhao2024,Ruske2021}.

Our model predicts the onset of the spontaneous flow and the interfacial instabilities occur concurrently at the same critical radius. Still, we find experimentally that the active fluid inside a stable droplet flows spontaneously without deforming the interface. Past experimental work also found that active fluids can spontaneously circulate when confined by solid walls~\cite{Wu2017} or in droplets with surface tension high enough to prevent deformation~\cite{Wioland2013, Alam2024, Chen2021}. We only consider the contribution of the active flows normal to the interface, and therefore our model does not address spontaneous flows that do not deform the interface. Another signature of the coupling of the interfacial and flow instabilities is that the characteristic length scale of the flow depends on surface tension. However, past work on the spontaneous flow instability for an active fluid confined between two parallel plates depends on the channel geometry \cite{Chandrakar2020} (but not surface tension because there is no interface present). To better understand the connection between these two instabilities, future work will focus on solving the droplet stability problem in 3D without invoking the lubrication approximation.

\textit{Acknowledgements:} We thank Thomas Videb{\ae}k and Aparna Baskaran for helpful discussions. We also thank Ray Adkins and Zvonimir Dogic for sharing the experimental protocols. This work was supported in part by the National Science Foundation through Grant no. MRSEC DMR-2011846 (BCS, GD, TRP), NSF CAREER award No. DMR-2047119 (GD) and CBET-2227361 (TRP, RAP). This research was also supported in part by grant NSF PHY-2309135 to the Kavli Institute for Theoretical Physics (KITP) (GD, TRP). We also acknowledge the use of the optical, microfluidics, and biomaterial facilities supported by NSF MRSEC Grant No. DMR-2011846. We acknowledge the support of the Natural Sciences and Engineering Research Council of Canada (NSERC), [567961-2022] (FC).
Cette recherche a été financée par le Conseil de recherches en sciences naturelles et en génie du Canada (CRSNG), [567961-2022] (FC).

\bibliography{refs}

\end{document}